\documentclass[]{aastex631}

\usepackage{ragged2e}
\usepackage{bm}

\submitjournal{ApJ}

\shortauthors{Lugaz et al.}
\graphicspath{{./}{figures/}}

\begin{document}

\title{A Coronal Mass Ejection and Magnetic Ejecta Observed In Situ by STEREO-A and Wind at 55$^\circ$ Angular Separation}

\correspondingauthor{No\'{e} Lugaz}
\email{noe.lugaz@unh.edu}

\author[0000-0002-1890-6156]{No\'{e} Lugaz}
\affiliation{Institute for the Study of Earth, Oceans, and Space, University of New Hampshire, Durham, NH, USA}

\author[0000-0001-6813-5671]{Tarik M. Salman}
\affiliation{Institute for the Study of Earth, Oceans, and Space, University of New Hampshire, Durham, NH, USA}

\author[0000-0002-5996-0693]{Bin Zhuang}
\affiliation{Institute for the Study of Earth, Oceans, and Space, University of New Hampshire, Durham, NH, USA}

\author[0000-0002-0973-2027]{Nada Al-Haddad}
\affiliation{Institute for the Study of Earth, Oceans, and Space, University of New Hampshire, Durham, NH, USA}

\author[0000-0002-5681-0526]{Camilla Scolini}
\affiliation{Institute for the Study of Earth, Oceans, and Space, University of New Hampshire, Durham, NH, USA}
\affiliation{CPAESS, University Corporation for Atmospheric Research, Boulder, CO, USA}

\author[0000-0002-2917-5993]{Charles~J. Farrugia}
\affiliation{Institute for the Study of Earth, Oceans, and Space, University of New Hampshire, Durham, NH, USA}

\author[0000-0002-2917-5993]{Wenyuan Yu}
\affiliation{Institute for the Study of Earth, Oceans, and Space, University of New Hampshire, Durham, NH, USA}

\author[0000-0002-9276-9487]{R\'eka~M. Winslow}
\affiliation{Institute for the Study of Earth, Oceans, and Space, University of New Hampshire, Durham, NH, USA}

\author[0000-0001-6868-4152]{Christian M\"ostl}
\affiliation{Space Research Institute, Austrian Academy of Sciences, Graz, Austria}

\author[0000-0001-9992-8471]{Emma~E. Davies}
\affiliation{Institute for the Study of Earth, Oceans, and Space, University of New Hampshire, Durham, NH, USA}

\author[0000-0003-3752-5700]{Antoinette~B. Galvin}
\affiliation{Institute for the Study of Earth, Oceans, and Space, University of New Hampshire, Durham, NH, USA}

\begin{abstract}

We present an analysis of {\it in situ} and remote-sensing measurements of a coronal mass ejection (CME) that erupted on 2021 February 20 and impacted both the Solar TErrestrial RElations Observatory (STEREO)-A and the {\it Wind} spacecraft, which were separated longitudinally by 55$^\circ$.  Measurements on 2021 February 24 at both spacecraft are consistent with the passage of a magnetic ejecta (ME), making this one of the widest reported multi-spacecraft ME detections. The CME is associated with a low-inclined and wide filament eruption from the Sun's southern hemisphere, which propagates between STEREO-A and {\it Wind} around E34. At STEREO-A, the measurements indicate the passage of a moderately fast ($\sim 425$~km\,s$^{-1}$) shock-driving ME, occurring 2--3 days after the end of a high speed stream (HSS). At {\it Wind}, the measurements show a faster ($\sim 490$~km\,s$^{-1}$) and much shorter ME, not preceded by a shock nor a sheath, and occurring inside the back portion of the HSS. The ME orientation measured at both spacecraft is consistent with a passage close to the legs of a curved flux rope. The short duration of the ME observed at {\it Wind} and the difference in the suprathermal electron pitch-angle data between the two spacecraft are the only results that do not satisfy common expectations. We discuss the consequence of these measurements on our understanding of the CME shape and extent and the lack of clear signatures of the interaction between the CME and the HSS.

\end{abstract}


\section{Introduction} \label{sec:intro}

The magnetic structure of coronal mass ejections (CMEs) is primarily known from direct {\it in situ} measurements of their magnetic field in the heliosphere and through reconstruction and fitting of magnetic field measurements, both in the photosphere and heliosphere \citep[e.g., see recent review by][]{Zhang:2021}. Multi-spacecraft measurements of CMEs, while rare, have been central in revealing that CMEs are global structures \citep[]{Burlaga:1982} that can often be understood as near-force-free flux ropes \citep[]{Burlaga:1988}. The Solar Terrestrial Relations Observatory \citep[STEREO;][]{Kaiser:2005}, composed of two nearly identical observatories, one ahead of Earth in its orbit (STEREO-A) and the other trailing behind (STEREO-B) was launched in 2006. In addition to an extensive suite of remote-sensing observations from the Sun Earth Connection Coronal and Heliospheric Investigation \citep[SECCHI;][]{Howard:2008} instruments, the mission promised to advance our understanding of CMEs by making multi-spacecraft {\it in situ} measurements of CMEs from two or three vantage points (including the Advance Composition Explorer (ACE) and {\it Wind} at L1). Because STEREO was launched during solar minimum, and each spacecraft separates from Earth by 22$^\circ$ per year, there were in fact only few multi-spacecraft measurements of CMEs by STEREO during its prime mission. These were summarized in \citet{Kilpua:2011}, with two main CMEs in May 2007 being extensively studied \citep[e.g., see][]{Liu:2008b,Moestl:2009a} as well as a CME in November 2007 when the two STEREO spacecraft were about 40$^\circ$ apart \citep[]{Farrugia:2011, Ruffenach:2012}. Other events in 2007--2008 as summarized by \citet{Kilpua:2011} were only observed by one of the two STEREO spacecraft in addition to L1, highlighting the fact that at longitudinal separations greater than 30--40$^\circ$, multi-spacecraft measurements of CMEs might be extremely rare. Consistent with this, \citet{Kilpua:2011} also discussed clear CME measurements at STEREO-A or STEREO-B in May--June 2008 that did not have associated measurements at L1 even though the separation between the STEREO spacecraft and the Sun-Earth line was only 25--30$^\circ$.

During the rising, maximum, and declining phases of solar cycle (SC) 24 (2010-2016), a variety of planetary missions with magnetometer measurements has made it possible to investigate the radial evolution of CMEs through conjunction events \citep[e.g.][]{Moestl:2015,Winslow:2015,Good:2015,Good:2016,Winslow:2016,Wang:2018,Davies:2020,Winslow:2021two,Salman:2020one,Lugaz:2020two,Palmerio:2021}. These works built upon similar efforts using Helios measurements during SC21 \citep[]{Bothmer:1998,Liu:2005,Leitner:2007} as well as work with Pioneer Venus Orbiter and NEAR in SC22 and the beginning of SC23 \citep[]{Mulligan:1999}. Taken together, this has made it possible to constrain the longitudinal extent of magnetic ejecta (ME) within CMEs to be typically around 20--30$^\circ$ \citep[]{Good:2016}, by investigating instances where a pair of spacecraft at different radial and longitudinal separations were able, or not, to measure the same CME.

However, analyzing data from spacecraft at different radial distances make it impossible to distinguish between changes in the magnetic structure due to the CME radial evolution and inherent deviation from a force-free or flux rope (FR) model. In fact, the evolution of CMEs through interplanetary space is significantly influenced by the heliospheric environment \citep[]{Temmer:2011,Manchester:2017}. This is primarily associated with three dynamic processes: (i) expansion, both in the radial and lateral direction  \citep[e.g.][]{Demoulin:2009,Gulisano:2010,Lugaz:2020two,AlHaddad:2021}, (ii) interactions with the structured background solar wind, such as corotating interaction regions (CIRs), high speed streams (HSSs), heliospheric current sheets (HCSs), and other CMEs \citep[e.g.][]{Winslow:2016,Lugaz:2017b,Liu:2019,Scolini:2020,Winslow:2021one,Palmerio:2021}, and (iii) the formation of the CME sheath region \citep[]{Siscoe:2008,Salman:2020two}.

There are clear cases of the same CME being observed by two spacecraft with separation greater than 30$^{\circ}$ but few for separations greater than 40$^\circ$ \citep[]{Good:2016}. In that work, the authors estimated that the probability for two spacecraft separated by 45--60$^\circ$ to measure the same ME  was about 10$\%$ and was 0\% for separations beyond 60$^\circ$. Their supplementary information includes one event (on 2011 November 17--20) that they identified as being observed by both MESSENGER and Venus Express while separated by 48.9$^\circ$ and 0.35~au, which is the largest angular separation between two spacecraft measuring the same ME. However, the event was not observed as a ME but only a shock by STEREO-B, which was positioned near 1~au in-between MESSENGER and Venus Express. \citet{Cane:1997} discuss Helios observations of CMEs, finding one event measured by two spacecraft separated by 53$^\circ$, and only another one for separations greater than 40$^\circ$. They also note that `` in a number of cases, two spacecraft were separated by less than 40$^\circ$, but an ejecta was seen at only one spacecraft.''
For the well-studied 2007 November CME which was observed by both STEREO spacecraft while separated by 40$^\circ$ \citep[e.g., see][]{Farrugia:2011}, the presence of the {\it Wind} spacecraft at L1 in-between the two STEREO spacecraft was critical in confirming that the same event was observed by STEREO-A and STEREO-B as the measurements by STEREO-A were strongly influenced by a fast stream which was already visible at L1. Some researchers \citep[]{Thoward:2009} in fact concluded that STEREO-A did not measure this event. 

However, longitudinal separations as small as 1$^{\circ}$ can also give rise to notable variance between measurements from one observing spacecraft to another \citep[e.g.][]{Lugaz:2018}. In that work, the authors pointed to the need for more investigations of multi-spacecraft measurements of CMEs at the same radial distance, as will be made possible by the return of STEREO-A to the proximity of the Sun-Earth line in 2022--2023. At present, we are in the ascending phase of SC25 with a new fleet of spacecraft in the inner heliosphere, including the Parker Solar Probe \citep[]{Fox:2016} and Solar Orbiter \citep[]{Mueller:2013} that open the way for more studies of radial conjunction between two or more spacecraft \citep[]{Winslow:2021two,Davies:2021}. In fact, the list of \citet{Moestl:2022} using data from spacecraft currently in the inner heliosphere, includes several CME events potentially measured {\it in situ} by two spacecraft. 

In this paper, we analyze a CME measured {\it in situ} by STEREO-A and {\it Wind} at L1 as the two spacecraft were separated by more than 55$^\circ$. This investigation comprises both the first CME since the ``return'' of STEREO-A to the vicinity of the Sun-Earth line in SC25 as well as the most distant multi-spacecraft measurement of a CME from two spacecraft at approximately the same radial distance. It therefore sheds light on the angular extent of MEs as they reach 1~au but also the variation in the morphology and property of the CME for separations of more than 30$^\circ$.

The rest of the paper is organized as follows. In section~\ref{sec:event}, we first present a general overview of the {\it in situ} measurements of the events before discussing all Earth-facing activity on the Sun to ensure that the same CME impacted both Earth/L1 and STEREO-A. We then present the remote-sensing observations associated with this CME and the observation associated with a coronal hole that it interacted with. We analyze the coronagraphic observations to obtain the CME speed, direction, and angular extent. In section~\ref{sec:insitu}, we analyze in-depth the {\it in situ} measurements made by STEREO-A and {\it Wind} of the CME and the high speed solar wind stream. We also discuss the presence of the shock and sheath ahead of the ME and analyze in detail the orientation of the ME. In section~\ref{sec:conclusion}, we discuss the implication of our results for the morphology of CMEs and conclude.

\section{Overview and Remote-Sensing Observations} \label{sec:event}

\subsection{Spacecraft Locations on 2021 February 20-24 and Instrumentation}

The heliocentric distances of STEREO-A and {\it Wind} were 0.966~au and 0.989~au, respectively, on 2021 February 20. The longitudinal separation was 55.7$^{\circ}$ with a latitudinal separation of 0.1$^\circ$ when measured in solar ecliptic coordinates, but $\sim 4.4^\circ$ when measured in Stonyhurst coordinates ({\it i.e}.\ where $z$ is the direction of the solar rotation axis). For this study, we rely primarily on STEREO-A/COR2 and EUVI \citep[]{Howard:2008}, LASCO/C2 and C3 coronagraphs \citep[]{Brueckner:1995}, and the Solar Dynamics Observatory (SDO)/Atmospheric Imaging Assembly (AIA) \citep[]{Lemen:2012} for remote-sensing observations, as well as {\it Wind}/3DP \citep[]{Lin:1995} and MFI \citep[]{Lepping:1995} and STEREO-A/PLASTIC \citep[]{Galvin:2008} and IMPACT \citep[]{Luhmann:2008} for {\it in situ} measurements. 

\subsection{Overview of the In Situ Measurements}

\begin{figure*}
\centering
{\includegraphics[width=1\hsize]{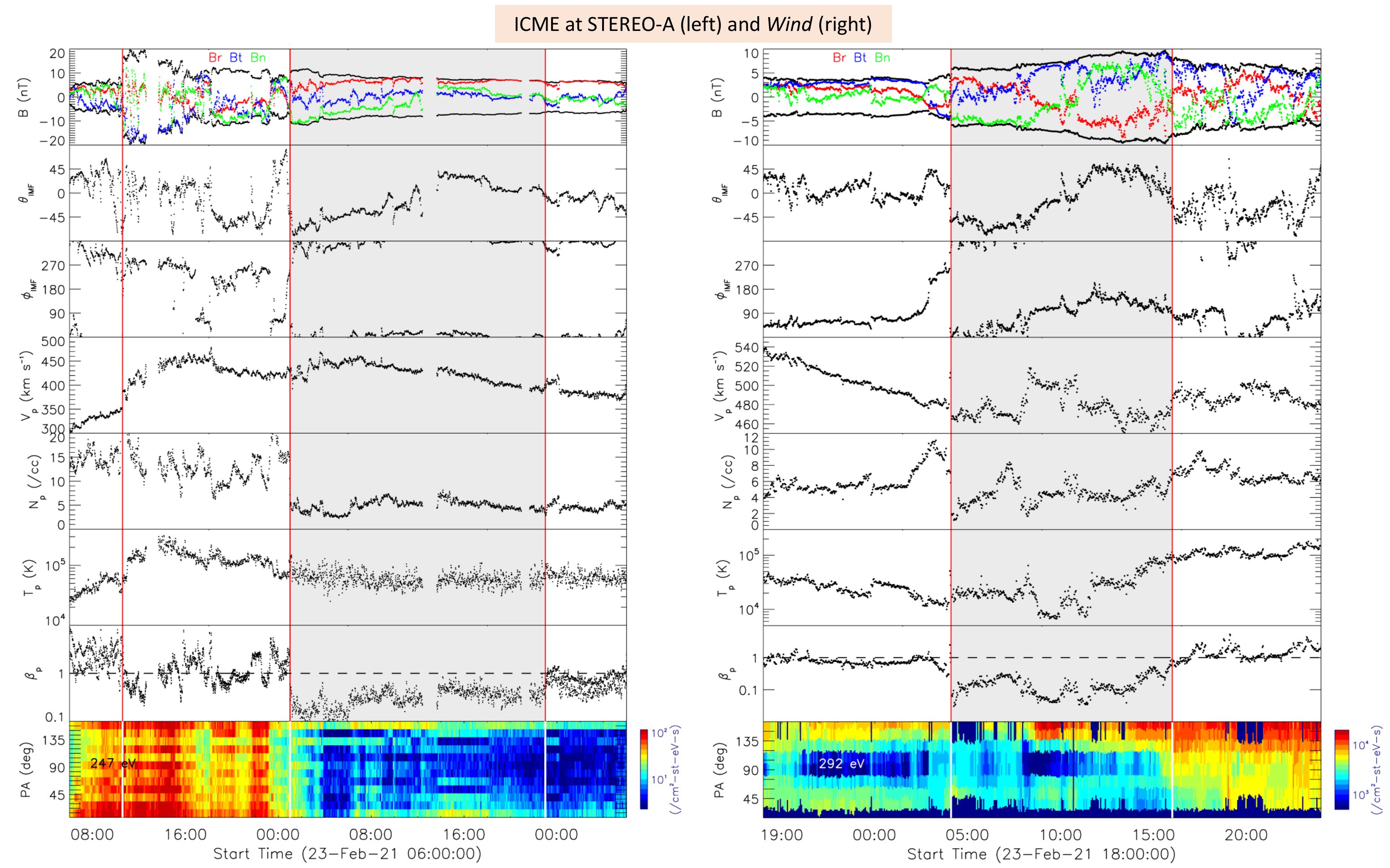}}
\caption{Overview of the CME measurements at STEREO-A (left) and {\it Wind} (right). The panels show from top to bottom, the total magnetic field, the radial, tangential and normal components of the magnetic field in RTN coordinates, the longitude and latitude of the magnetic field angle, the proton velocity, density, temperature and $\beta$ and the pitch-angle distribution of suprathermal electrons at about 250 eV (247 eV channel at STEREO-A and 292 eV channel at {\it Wind}). The vertical lines show the shock (at STEREO-A), and start and end times of the ME (at both spacecraft), which is also highlighted in grey.} 
\label{fig:figure1}
\end{figure*}

Figure~\ref{fig:figure1} shows plasma and magnetic field measurements of a CME at STEREO-A and {\it Wind} on 2021 February 23--24. A CME preceded by a fast-forward shock impacts STEREO-A at 10:34~UT on February 23. There is a clear sheath region characterized by hot, magnetized, and turbulent plasma that stops around 01:00~UT on February 24 with the start of the ME. At this time, there is a clear drop in density, increase in magnetic field, and decrease in the magnetic field variability and a period of low proton $\beta$. As is relatively common, the end time of the ME is not clear as the magnetic field strength slowly decreases to its pre-event value. We pick an end time of 23:00~UT on February 24 corresponding to a small increase in velocity and density. At this time, the magnetic field is almost purely radial and down to less than 7~nT. The ME has an average speed of 426 km\,s$^{-1}$.

At {\it Wind}, the leading edge of a short ME, embedded in the declining part of a HSS arrives at 04:09~UT on February 24. The ME at {\it Wind} is not preceded by a sheath and a fast-forward shock, as is the case for the STEREO-A measurements. It is however clearly a magnetically dominated (low proton $\beta$) structure, with enhanced magnetic field strength and relatively smooth rotation of the magnetic field vector. The end time of the ME is again not clear but we choose February 24 at 16~UT as a likely end time. This is based on the reversal of the $B_N$ and $B_R$ components of the magnetic field as well as the gradual increase in velocity, density, temperature and proton $\beta$. The ME has an average speed of 490 km\,s$^{-1}$.

Based on the {\it in situ} measurements, this is possibly the same CME being measured at both spacecraft, a conclusion reached in the database of \citet{Moestl:2022}, who additionally support the connection with heliospheric imaging by STEREO-A/SECCHI. In particular, both spacecraft measure a south-to-north rotation of the magnetic field ($B_N$ from negative to positive) and the start times of the ME are only $\sim$ 3 hours apart at both spacecraft. In the next section, we discuss surface, coronal and heliospheric imaging observations of CMEs during this time interval to confirm that the same CME is indeed observed at both spacecraft.

\subsection{CME Eruption Candidate}

We use the CDAW CME catalog \citep[]{Yashiro:2004} as well as visual inspection of EUV and coronagraphic observations to determine all eruptions that could impact STEREO-A and/or spacecraft at L1 point in 2021 February 24. To do so, we focus on eruptions that occur between 2021 February 17 and 21, corresponding to an average propagation speed of $\sim$250 to 850 km\,s$^{-1}$ for a CME arrival on early February 24. Due to the positioning of STEREO-A, CMEs that impact Earth should appear as (partial) front-sided halo CMEs for LASCO observations and near-western limb event for STEREO-A/COR2 observations, while CMEs that impact STEREO-A should appear as (partial) front-sided halo CMEs for STEREO-A/COR2 observations and near-eastern limb event for LASCO observations. We note that, in this section, we do not require a single CME hitting STEREO-A and the spacecraft at L1 point simultaneously. Here, we find three eruption candidates that might account for the {\it in situ} measurements at one or both of the spacecraft.

The first candidate is a CME with first appearance time in LASCO C2 field of view (FOV) at 23:12 UT on February 17. The associated filament eruption is visible in STEREO-A EUVI 304 $\mathring{\rm{A}}$, and this filament is roughly facing STEREO-A in longitude (not shown here). This CME is relatively poor and only visible in coronagraph observations by LASCO. Furthermore, the propagation direction of this CME is out of the ecliptic plane ($\sim$30S in LASCO C3 FOV), and its angular width is quite small ($\sim$25$^\circ$ based on the CDAW catalog), which may together indicate the unlikelihood of the CME hitting STEREO-A. In addition, there are no clear measurements in C3 beyond about 10~$R_\odot$ and the speed is greater than 550 km\,s$^{-1}$ making it unlikely to match the arrival at STEREO-A.

The second candidate is the CME with the first appearance time in LASCO C2 FOV at 11:24 UT on February 20. According to the CDAW catalog, this CME is intermediate fast with a speed $\sim$700 km\,s$^{-1}$ (the leading edge speed at 20 $R_\odot$ derived from the quadratic fit), and has a central position angle of $90^\circ$ and angular width of $207^\circ$ in LASCO FOV. We find that this CME is the only major eruption which could possibly hit STEREO-A and {\it Wind}, and the details of the related remote-sensing observations are shown in Section \ref{sec_remote_obs}. It is the only clear partial halo in that time period. It also appears as a partial halo from the western limb in STEREO-A/COR2 with a first image at 12:53~UT on 2021 February 20. From the images, it is clear that the CME propagates between the Sun-Earth line and the Sun-STEREO-A line.

The last candidate is only visible by SDO/AIA images associated with a very weak filament eruption on February 19 from the northern hemisphere and close to disk center as seen from Earth. However, there is no CME counterpart in coronagraph including from STEREO-A where it should appear as a relatively clear western limb event. As such, we consider this might be a failed filament eruption.
Overall, the only {\it in situ} measurements recorded by STEREO-A and/or Wind during this time period are those on February 24 as described above. Hereafter, we focus on the 2021 February 20 CME as the eruption that impacted both STEREO-A and L1 on February 24.

\subsection{Surface and Coronal Observations of the 2021 February 20 CME}\label{sec_remote_obs}

\begin{figure}[!hbt]
	\centering
	\includegraphics[width=\textwidth]{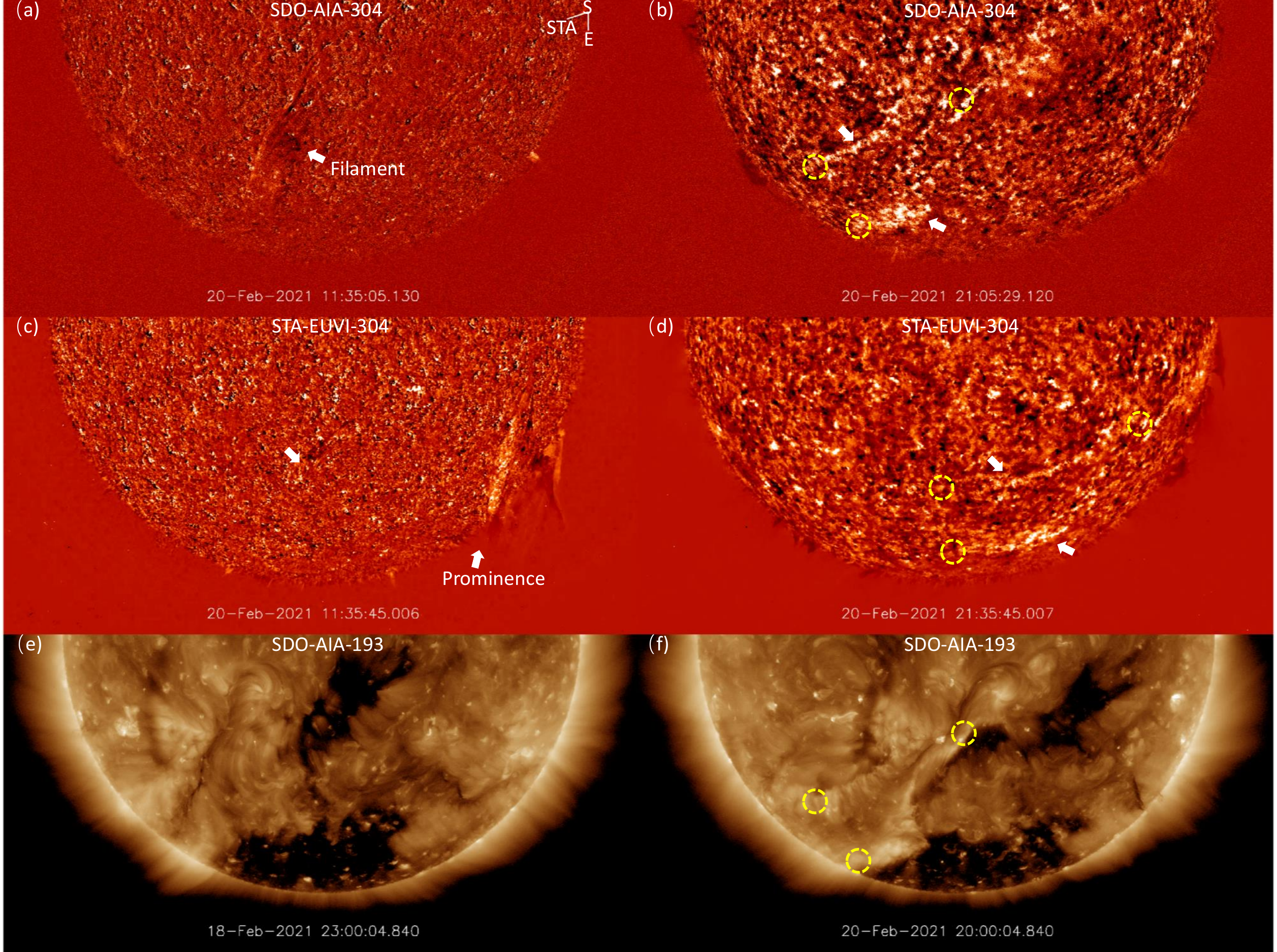}
	\caption{\small Running-difference (a) and base-difference (b) images from SDO AIA 304 $\mathring{\rm{A}}$. Running-difference (c) and base-difference (d) images from STEREO-A EUVI 304 $\mathring{\rm{A}}$. (e)-(f) Image from SDO AIA 193 $\mathring{\rm{A}}$. Note that every panel is shown at different time step and by different image-processing methods to enhance the visibility of different structures. The insert in panel (a) shows the locations of STEREO-A and Earth relative to the Sun. The arrow in panels (a) and (c) show the filament/prominence. The arrows in panels (b) and (d) show the ribbons with maximum extent shown with yellow circles. The circles in panel (f) indicate the same locations as determined by the circles in panel (b). An animated version of the Figure shows first, a two hour and forty minute animated version of panel (c) highlighting the deflection of the prominence, then a seven-hour animated version of panel (d) highlighting the formation of the ribbons, and finally a five-day version of panels (e) and (f) highlighting both the long duration coronal hole and the opening of new magnetic flux in its vicinity following the eruption.} 
	\label{euvi_aia}
\end{figure}

Figure \ref{euvi_aia} shows the observations of the signatures of the 2021 February 20 CME by STEREO-A EUVI and SDO AIA both in 304 $\mathring{\rm{A}}$ wavelength at different time steps. To enhance the visibility of the desired structure, we show the images at different time steps and by different image-processing methods. This CME is associated with a filament/prominence (the same structure but observed from different view angles; hereafter we use the term filament only) eruption from a solar quiet region in the southern hemisphere (Figures \ref{euvi_aia}(a) and (c)). After the eruption of the filament, it experiences a northward deflection observed by STEREO-A EUVI as seen in a movie version available in the online article. Such a northward deflection would ensure this CME impacts spacecraft in the ecliptic plane. In Figures \ref{euvi_aia}(b) and (d), the two bright ribbons after the CME eruption are observed (indicated by the white arrows), and last discernible for more than 12 hours until about 21:30 UT. The easternmost and westernmost locations of the bright ribbons are shown by the yellow circles in Figures \ref{euvi_aia}(b) and (d). The two bright ribbons indicate the locations of the CME footpoints on the solar surface, which is found to roughly extend $\sim$ 40$^\circ$ in latitude and $\sim$ 60$^\circ$ in longitude. It indicates that the CME has a wide extent, especially in longitude, i.e., the angular width is large from a polar viewpoint. 
Furthermore, comparison between the latitudinal extension and longitudinal extension may indicate that the axis of the eruptive CME flux rope structure is low inclined with respect to the solar equator. Those surface observations are consistent with the derived CME propagation parameters as described below.

\begin{figure}[!hbt]
	\centering
	\includegraphics[width=0.8\textwidth]{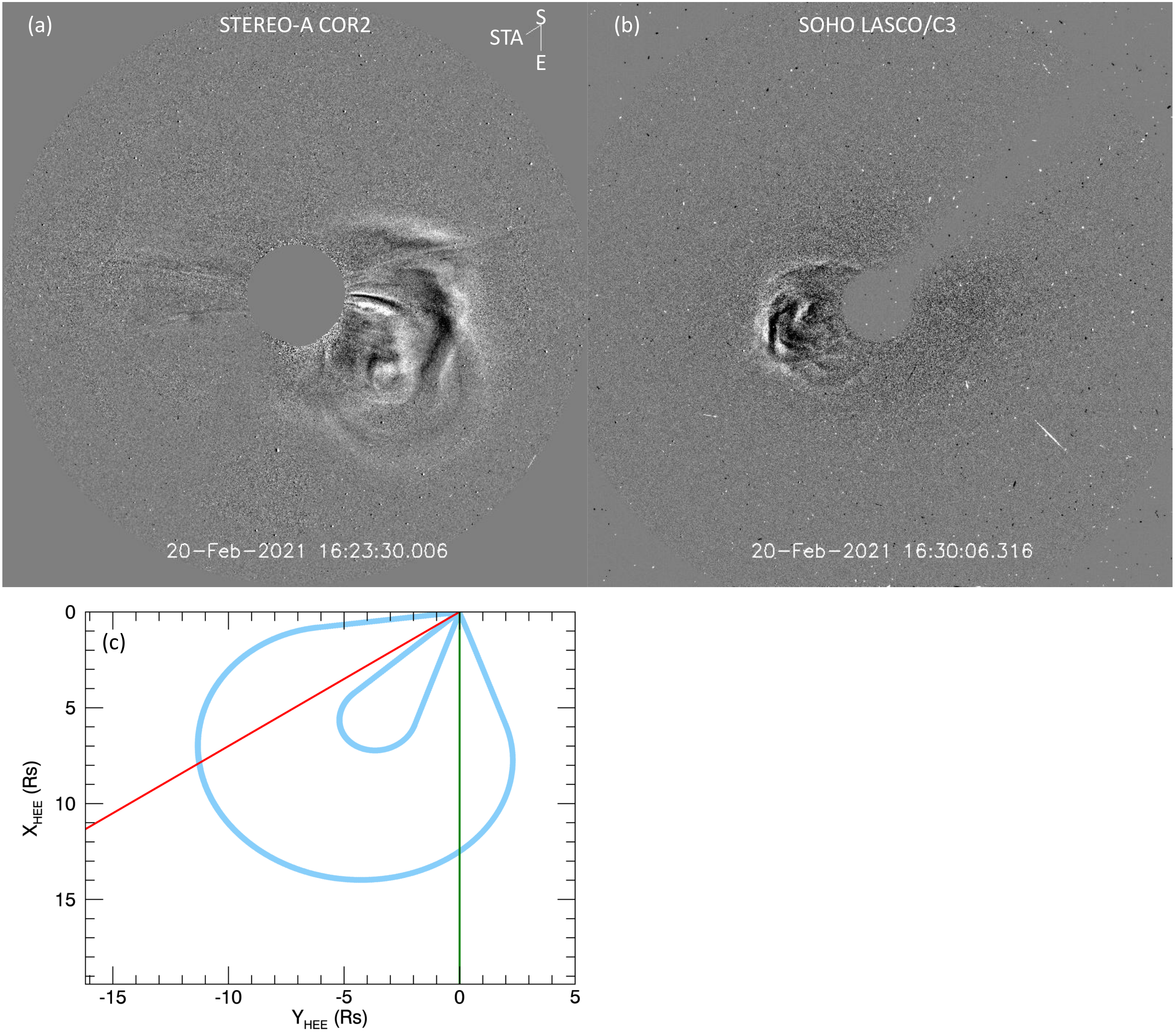}
	\caption{\small (a)--(b) Running-difference images by STEREO-A COR2 and LASCO C3 at roughly the same time. (c) Intersection of the reconstructed flux rope using the GCS model in the ecliptic plane. The red and green lines show the Sun-STEREO-A and Sun-{\it Wind} lines, respectively.}
	\label{coronagraph}
\end{figure}

The CME is then observed by coronagraphs on board STEREO and LASCO. Figures \ref{coronagraph}(a) and (b) show the running-difference images of the CME in STEREO COR2 and LASCO C3 at roughly the same time. It is found that this CME appears as partial halo in both coronagraph images. 
To obtain the CME propagation parameters in the 3-dimensional (3D) space, we use the graduated cylindrical shell (GCS) model \citep{2006ApJ...652..763T,2009SoPh..256..111T} which assumes that the CME has a flux rope structure and self-similar expansion in the corona. 
Although the CME experiences a deflection in EUVI FOV, there is no significant deflection in STEREO-A COR2 and LASCO C3 FOV. We perform the GCS model fitting at different time steps by only changing the height and maintaining the other free parameters. The CME speed is estimated by linearly fitting the height-time measurements. The CME 3D propagation parameters derived by the GCS model are direction of $\theta=-22^\circ$ in latitude, $\phi=-34^\circ$ in longitude, tilt $\gamma=-16^\circ$, an average speed of $v=714$~km\,s$^{-1}$, face-on width $w_f=106^\circ$, and edge-on width of $w_e=48^\circ$. This indicates a low inclination angle with respect to the solar equator and a relatively large face-on angular width. \citet{Lee:2015} performed an investigation of the properties of 44 halo CMEs using the GCS reconstruction technique and found that only 9 of the 44 studied events had a face-on width greater than 100$^\circ$.
In order to find out how the CME crosses STEREO-A and {\it Wind}, we show the intersection flux rope structure in the ecliptic plane in Figure \ref{coronagraph}(c). The red line shows the Sun-STEREO-A line, and the green one shows the Sun-{\it Wind} line. It is found that this CME can hit both STEREO-A and {\it  Wind} given its initial width and direction. 

\subsection{Earth-facing Coronal Hole}\label{sec:CH}

There is a clear equatorial coronal hole in SDO/AIA images that lies in the southern hemisphere and crosses the central meridian around 20~UT on February 18 (see Figure \ref{euvi_aia}, panel (e)).
This coronal hole is the source of the high speed solar wind stream (HSS) observed in {\it in situ} by {\it Wind} ahead of the arrival of the CME. The interaction of the CME with this HSS is discussed with further details below.
The eruption of February 20 results in dimming regions or opening of new field south-east of the coronal hole, which persists until it rotates out of the SDO field-of-view. The newly ``opened'' magnetic field (transient coronal hole) associated with the CME eruption is Earth-facing at the eastern end of the coronal hole in Figure \ref{euvi_aia} (f), marked by the western most yellow circle.
The three yellow circles in that panel mark the same locations of the bright ribbon maximum extent as determined by the circles in Figure \ref{euvi_aia}(b). 

\subsection{Heliospheric Propagation: Drag-Based Modeling}

While the CME is clearly observed in STEREO-A/HI \citep[e.g., see][]{Moestl:2022}, we focus here primarily on the drag-based modeling (DBM) of its transit. In general, the results from single-spacecraft fitting of the CME leading edge based on the STEREO-A heliospheric imager (HIA) data of $\phi_{HIA}=-26^\circ$ longitude are consistent with the GCS direction, and the average speed of the CME nose in the STEREO-A/HI field of $v=423$~km\,s$^{-1}$ points to a clear deceleration as the CME propagates into interplanetary space \citep{Moestl:2022}.

However, any information about the CME kinematics comes with relatively significant uncertainties due to the single viewpoint with the lack of STEREO-B and the direction of propagation close to the Sun-STEREO-A line, making any physical deceleration harder to distinguish from the apparent acceleration \citep[]{Lugaz:2013a}. In particular, as the CME experiences deceleration, the true direction shall be closer to the Sun-STEREO-A line and the speed faster than what is derived by the single-spacecraft fitting method.

We use the DBM formulated by \citet{Vrsnak:2013} to estimate the CME arrival times (front boundary of the magnetic ejecta) and impact speeds at STEREO-A and L1 for consistency. While the DBM is typically used to investigate the arrival of the shock/sheath, here we focus on the arrival of the front of the magnetic ejecta to be consistent between STEREO-A and L1. The DBM solves for the CME kinematics under the assumption that the drag is the only force acting on CMEs in the heliosphere. We use it here to confirm that this CME is able to impact both spacecraft with approximately the measured speed. We use the ``advanced'' version of the DBM, which takes into consideration the direction of the CME, the angular separation with the measuring spacecraft, and assumes a self-similar cone-like CME.

We use an initial speed of 710~km\,s$^{-1}$, a time at 20~$R_\odot$ of 19:45~UT on February 20, a direction of E34 and a half-angle of 45$^\circ$ based on the GCS reconstruction as inputs into the DBM. We then determine the CME (front boundary of the CME or the ME) arrival time at STEREO-A and {\it Wind}. We use the measured solar wind speed upstream of the CME of 325~km\,s$^{-1}$ at STEREO-A and 500~km\,s$^{-1}$ at {\it Wind}. We adjust the drag parameters to approximately match the CME impact speed of $\sim$ 420~km\,s$^{-1}$ at STEREO-A and $\sim$ 480~km\,s$^{-1}$ at {\it Wind} as well as the arrival time. For a drag parameter of 0.22 $\times$ 10$^{-7}$ km$^{-1}$, the CME arrival time at STEREO-A derived by the DBM is 22:05~UT on February 23, as compared to the 01:00~UT ME start time on February 24, based on {\it in situ} magnetic field and plasma signatures (the  speed forecasted by the DBM is 412~km\,s$^{-1}$). For a drag parameter of 0.4 $\times$ 10$^{-7}$ km$^{-1}$, the derived ME arrival time is 22:45~UT on February 23 at {\it Wind} with a forecasted speed of 468~km\,s$^{-1}$. While this is not in perfect agreement with the 04:09~UT ME start time on February 24, this is still a relatively decent match. This is the case even though the CME is preceded by a shock at STEREO-A and not at {\it Wind}. Furthermore, we note that the values of the drag parameter used here are consistent with past studies. \citet{Vrsnak:2013} for example found that $\gamma$ is in the range of 0.2 -- 2 $\times 10^{-7}$~km$^{-1}$ based on a statistical analysis of CME transit times as well as an analysis of the parameters used to derive $\gamma$, while \citet{Calogovic:2021} found that $\gamma = 0.3 \times 10^{-7}$ km$^{-1}$ was the optimal fixed value for a set of 146 CMEs studied via an ensemble version of the drag-based model.  
\section{In situ Measurements} \label{sec:insitu}

\subsection{Shock and Sheath Measurements at STEREO-A}

A fast-forward shock driven by the ME is measured at STEREO-A at 10:34~UT on February 23. It is propagating in a slow solar wind with speed of 325~km\,s$^{-1}$ and fast-magnetosonic speed of 37~km\,s$^{-1}$. We perform a Rankine-Hugoniot analysis of the shock parameters. The upstream and downstream states are determined over an interval of eight minutes each, in a way not to include the shock ramp, similar to \citet{Kilpua:2015a}. The measured upstream-to-downstream jumps for the magnetic field and proton density are $\sim$2.8 and $\sim$1.8 respectively. The solar wind speed jump across the shock ramp is measured to be $\sim$36~km\,s$^{-1}$.

We then estimate the shock normal direction using the magnetic coplanarity. At STEREO-A, the shock normal is estimated to be (0.83 $\pm 0.01$, $-0.10$ $\pm 0.03$, $-0.54$ $\pm 0.02$) in $RTN$ coordinates. The error bars are estimated using slightly different upstream and downstream intervals (1 minute before and after the chosen interval with the error representing the full range of variation). The shock normal angle (angle between the shock normal direction and the upstream magnetic field) is found to be $\sim$62$^\circ$, corresponding to a quasi-perpendicular shock at STEREO-A. The angle between the shock normal and radial direction, which can be used as an approximation of a spacecraft crossing distance from the CME nose \citep[e.g.][]{Paulson:2012,Janvier:2015} is $\sim$34 $\pm 1^\circ$. This can be used as an argument that the spacecraft crossing for STEREO-A occurs away from the nose of the shock. The shock speed in the spacecraft reference frame is $\sim$ 360~km\,s$^{-1}$, corresponding to a Mach number of 1.8. 

The STEREO-A spacecraft encountered the sheath for a period of 14.5 hours (see Figure~\ref{fig:figure1}, left). The sheath-to-ME duration ratio of $\sim$0.66 is significantly higher than a typical value of $\sim$0.31 \citep[]{Jian:2018, Salman:2020two}. The sheath thickness in the radial direction is $\sim$ 0.15~au. The sheath is $\sim$2.2 times more magnetized and $\sim$3.3 times hotter compared to the unperturbed solar wind upstream of the shock. 
However, the sheath density is comparable to the background solar wind, which contrasts with a typical CME sheath. Strong density compression is only observed in the vicinity of the shock ramp and not beyond. Except for the very front of the sheath, the sheath velocity profile is consistent with a roughly constant speed. From previous work, such a sheath is expected to be driven by a ME with a relatively weak magnetic field and moderate-fast leading edge speed in the solar wind frame \citep[]{Salman:2021}. This is consistent with the relatively weak ME with 8.2~nT average magnetic field strength and a M$_{pseudo}$ (the ``Mach'' number of the ME front in the solar wind frame) of 2.7 for the ME. In the density and proton beta profiles, the transition from the sheath to the ME is well-defined and clearly represents the start of the ME.

There is no shock nor sheath measured upstream of the ME at {\it Wind}. The upstream solar wind speed is about 500~km\,s$^{-1}$ and the upstream fast magnetosonic speed is about 47~km\,s$^{-1}$. The front of the ME has a similar speed to that of the solar wind and it is therefore consistent with the lack of shock.

\subsection{High Speed Stream at STEREO-A and L1}

\begin{figure*}
\centering
{\includegraphics[width=1\hsize]{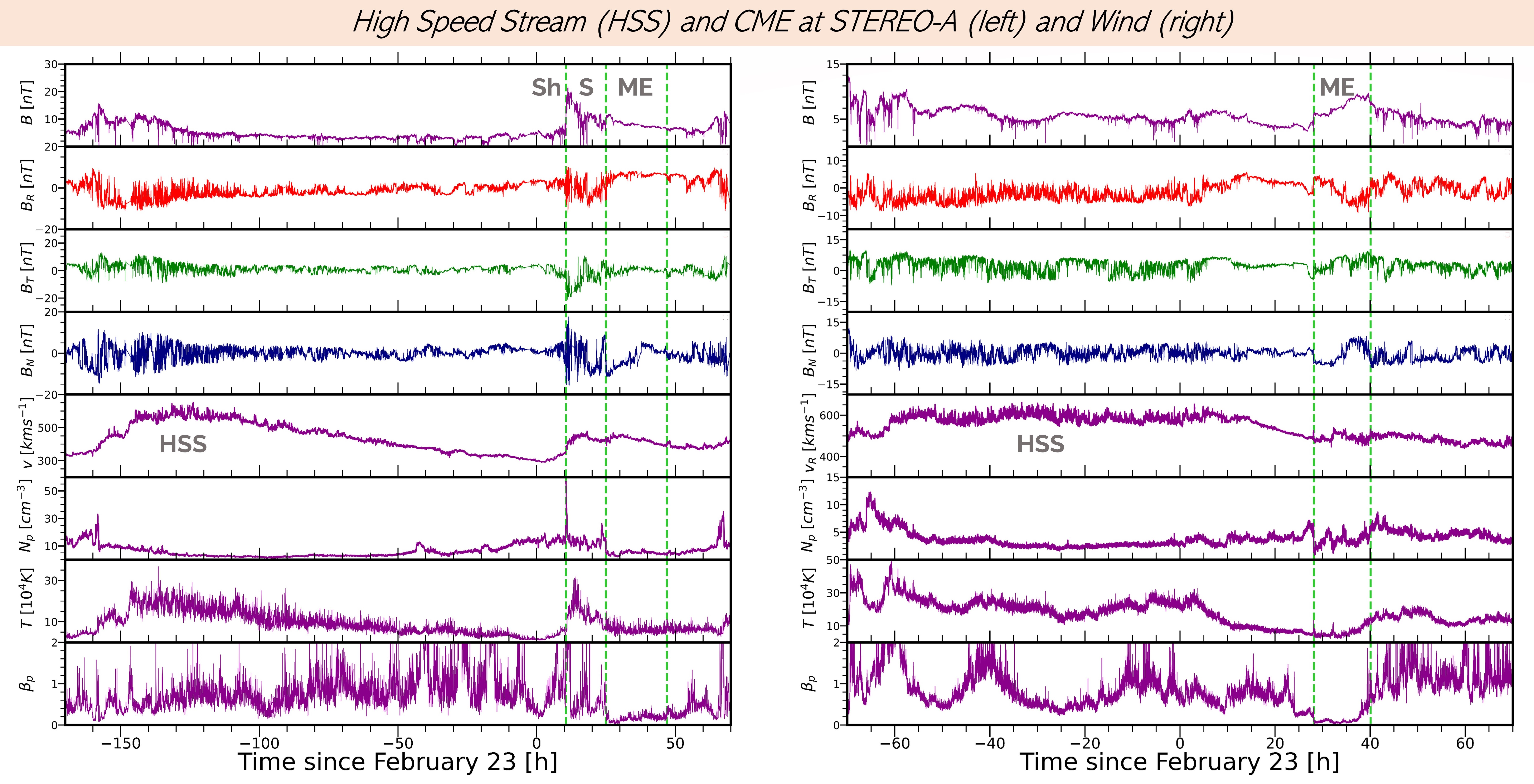}}
\caption{Measurements of the HSS and ME at STEREO-A (left) and {\it Wind} (right). The panels show from top to bottom, the total magnetic field, the radial, tangential and normal components of the magnetic field, the proton velocity, density, temperature and $\beta$. The vertical lines show the shock (at STEREO-A), and start and end times of the ME (at both spacecraft)} 
\label{fig:figure2}
\end{figure*}

At {\it Wind}, the ME is embedded at the back of a HSS as clearly seen in Figure~\ref{fig:figure2}. This HSS starts on February 20 and is associated with the coronal hole described in section~\ref{sec:CH}. There is a compression region and stream interface on February 20 and a period of $\sim$ 3 days with a solar wind speed of about 600~km\,s$^{-1}$. The ME starts at the back of the HSS after a $\sim$ 12-hour period of decreasing speed, low temperature and primarily radial outward field, probably associated with the rarefaction region behind the HSS. 

The HSS is clearly visible at STEREO-A on February 17--21 with a compression and stream interface on February 16. The expected corotation time from STEREO-A to L1 is about 3.8 days assuming a corotating rate of 14.5$^\circ$ per day  \citep[]{Jian:2019, Allen:2020}. This is consistent with the delay of $\sim 4.1--4.2$ days between the stream interface at STEREO-A and {\it Wind}. 
When the CME launches from the Sun (around 12~UT on February 20), STEREO-A is measuring the back of the HSS, with solar wind speed below 450~km\,s$^{-1}$. At the same time, {\it Wind} is still inside the interaction region with elevated magnetic field and density and the spacecraft remains inside the HSS from late on February 21 to around 12~UT on February 23, {\it i.e}.\ during most of the Sun-to-Earth propagation of the CME. It is however likely that the whole CME did not interact much with the HSS on its way to Earth due to the curvature of the Parker spiral. Most interaction would have happened relatively close to the Sun, probably around February 21 and continued only through the western leg which is closer to the HSS.

\subsection{Comparison of the ME at STEREO-A and L1}

At STEREO-A, the ME is measured to be slowly expanding, with an expansion speed of 16~km\,s$^{-1}$, corresponding to a dimensionless expansion parameter of $\zeta \sim 0.15 $ \citep[see][for a definition of $\zeta$]{Gulisano:2010}. The ME has a relatively weak magnetic field strength ($\sim 8$~nT) and low $\beta$. The average speed is 425~km\,s$^{-1}$. The crossing time of 22 hours corresponds to a measured size of 0.225~au, which is relatively typical. 

At {\it Wind}, the ME does not expand but has a relatively complex but overall flat speed profile. The ME also has a relatively weak magnetic field strengthn ($\sim 7.5$~nT) and low $\beta$. The average speed is 490~km\,s$^{-1}$, meaning it is actually faster than at STEREO-A but does not drive a shock. The crossing time of 12 hours corresponds to a measured size (diameter) of 0.14~au, which is small. Typical values for the ME size near 1~au are about 0.25 $\pm$ 0.12~au \citep[]{Bothmer:1998,Lepping:2006}, and \citet{Salman:2020two} found a similar size for MEs that do not drive a sheath. As such, the ME measured by {\it Wind} is on the lower end of ME sizes whereas it is about typical at STEREO-A. Overall, this indicates that the ME has been affected by the interaction with the HSS, at least through a faster upstream solar wind speed that hindered the formation of a shock or even of any sheath region. The flat speed profile inside the ME and the relatively short duration are consistent with some compression from the moderately fast solar wind behind the ME. The lack of any sheath signatures can be considered somewhat puzzling as many slow CMEs without shocks are still associated with sheath signatures such as density and magnetic compressions \citep[]{Salman:2020two},  and the same ME drives a shock and sheath at STEREO-A. We hypothesize that the presence of the HSS next to the western leg of the ME (the part of the ME impacting {\it Wind}) had two consequences: (1) this leg of the ME is mostly convected with the HSS, (2) the expansion of this part of the ME is hindered by the HSS. As such, this part of the ME is not able to drive any sheath.

As shown in Figure~\ref{fig:figure1}, the measurements at STEREO-A are consistent with the presence of bi-directional electrons (BDEs) throughout most of the ME (the exception being a period of about 2 hours from 4 to 6 UT). At {\it Wind}, there is a mix of BDEs (for about 40\% of the event, especially  between 8 and 11:30~UT but also for a short period around 6 UT) as well as what appears as single-strahl electrons in the 180$^\circ$ sector, indicating alternating open and closed field lines inside the ME. This is again consistent with the part of the ME impacting {\it Wind} to have been highly affected by the interaction with the HSS and interaction of that part of the ME (the western leg) with the open magnetic field lines at the Sun associated with the coronal hole.

We investigate the suprathermal measurements at {\it Wind} in more depth. There is a clear depletion of suprathermal electrons around pitch-angle (PA) 90$^\circ$ throughout the ME at {\it Wind}. During the times with unidirectional strahls, the intensities along PA 160--180$^\circ$ is about one order of magnitude larger than along PA 0--20$^\circ$, which are themselves comparable or slightly more elevated than those along PA 90$^\circ$. During these time periods (from 11:30 UT to 16 UT), the magnetic field $B_R$ component at {\it Wind} is negative (sunward), meaning that PA 180$^\circ$ represents electrons flowing outward from the western leg of the CME. These would travel much less distance before being detected at {\it Wind} than those with PA 0$^\circ$ coming from the eastern leg. It is therefore possible that those seemingly unidirectional strahl measurements correspond to closed but very asymmetric magnetic field lines \citep[a similar argument was made by][]{DeForest:2013}.  
Figure~\ref{euvi_aia} (f) clearly shows a dark region associated with a transient coronal hole on the southeast of the main coronal hole. This region is close to the westernmost ribbons shown in panel (b) at approximately the same time. From this, it is possible that the western leg of the CME is experiencing a large amount of interchange reconnection and contains a significant portion of open field lines.

\subsection{Orientation of the ME}

The orientation of the ME can be easily seen already in Figure~\ref{fig:figure1} with a south-to-north rotation ($B_N$ negative to positive) at both spacecraft and the component of the field in the ecliptic primarily in $R$ (anti-sunward) direction at STEREO-A and in the $-R, T$ direction at {\it Wind}.

We perform force-free fittings of the ME following \citet{Lepping:1990}. We do not fix the $\alpha$ parameter (related to the twist at the boundary) but make it one of the fitting parameters. At STEREO-A, the orientation of the ME axis is $\left(-4^\circ, 43^\circ\right)$ with an impact parameter of 0.54, $\alpha R = 2.07$ and a ME size of 0.185~au.  At {\it Wind}, the orientation of the ME axis is $\left(-4^\circ, 157^\circ\right)$ with an impact parameter of 0.41, $\alpha R = 2.61$ and a ME size of 0.058~au. At both spacecraft, the ME has a low inclination, and the orientation is also somewhat consistent with the visual inspection. The axial field is in the $R, T$ direction at STEREO-A and in the $-R, T$ direction at {\it Wind}. This is consistent with a low-inclined south-west-north (SWN) cloud which is crossed on its east leg at STEREO-A and west leg at {\it Wind}. The impact parameter at both spacecraft indicates a ME propagating south of the Sun-spacecraft plane, which is consistent with the remote-sensing imaging. 

Both crossings occur close to the legs. The $\lambda$ of \citet{Janvier:2015} which varies between $0^\circ$ at the ME nose and $\pm 90^\circ$ in the legs is 47$^\circ$ for STEREO-A and $-67^\circ$ for {\it Wind}, confirming leg crossings at both spacecraft ($\lambda$ is the angle between the axis of the ME and the ortho-radial direction). The $\lambda$ parameter for the ME at STEREO-A is larger by about 14$^\circ$ than the angle between the shock normal and the radial direction, indicating that the shock has a larger radius of curvature than the ME, consistent with common expectations. The fact that this is a leg crossing and the difference between these two angles are also consistent with the relative large sheath size as compared to the ME size at STEREO-A. 

One unexpected result is that, while the CME is crossed closed to the legs, the duration of the ME is not large, as would be expected from a twisted flux rope model. This could be an issue with the selection of the boundaries. At STEREO-A, there is in fact an additional period of low density and low $\beta$ with primarily radial magnetic field following the ME, which would be consistent with the crossing through a mostly untwisted leg. At {\it Wind}, the ME is quickly followed by a period of high $\beta$ and more complex magnetic field, i.e. not  consistent with the crossings through an untwisted ME leg.

\section{Discussion and Conclusions} \label{sec:conclusion}

\begin{figure*}
\centering
{\includegraphics[width=1\hsize]{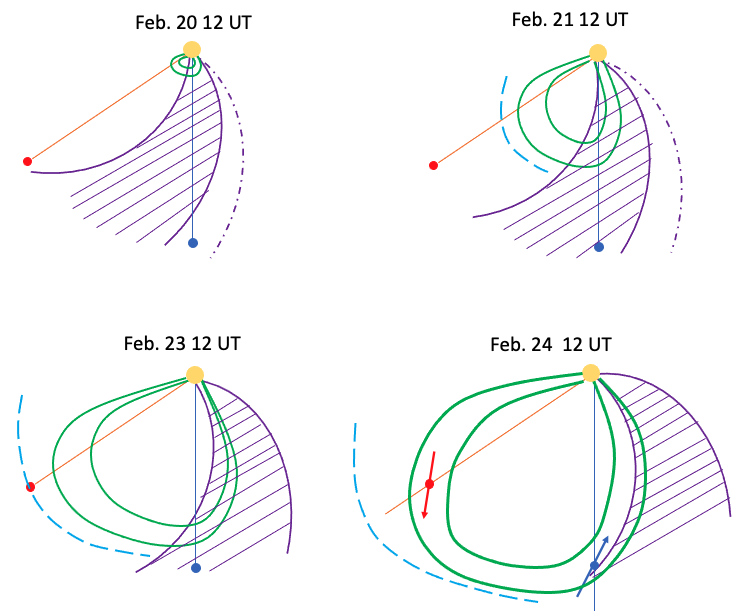}}
\caption{Sketch of the CME viewed from solar north as it propagates. The Sun is shown with the yellow disk, STEREO-A the red one and L1 the blue disk. The ME is drawn as a green flux rope shape. The HSS is shown in hatched purple, while the stream interaction region is in dashed-dotted purple. The ME-driven shock is in dashed light blue. In the last panel, the arrows indicate the reconstructed orientation of the ME at the two spacecraft.} 
\label{fig:sketch}
\end{figure*}

We have analyzed multi-spacecraft {\it in situ} measurements of a ME by STEREO-A and {\it Wind} in February 2021 as they were separated by $\sim 55^\circ$ in longitude. We identify the eruption that caused the ME as a filament eruption from the southern hemisphere of the Sun in close proximity to a coronal hole. Coronagraphic measurements indicate that the CME propagates in-between the Sun-STEREO-A and Sun-Earth line. The {\it in situ} measurements and force-free fitting are consistent at both spacecraft with crossings through the CME legs. At STEREO-A, the ME is slower than at {\it Wind} and it travels through typical slow solar wind and drives a shock and a sheath. BDEs indicate that the magnetic field lines inside the ME are mostly closed at STEREO-A. At {\it Wind}, the ME is embedded in the back of a fast solar wind stream and it does not drive a shock nor a sheath. The suprathermal electron measurements indicate a complex mix of magnetic topology throughout the ME. 

Overall, the picture from the coronagraphic observations, the presence of a coronal hole with a HSS and the joint {\it in situ} measurements at STEREO-A and {\it Wind} allow us to paint the sketches summarized in Figure~\ref{fig:sketch} and described in the scenario below.
As the CME erupts, it is almost entirely embedded inside the open magnetic field regions associated with a coronal hole (top left panel). At the time of the event, the associated HSS is passing STEREO-A and the stream interaction region associated with it is at {\it Wind}.   On February 21, the HSS associated with the coronal hole impacts {\it Wind}. At this time, the western leg of the CME is fully embedded inside the HSS whereas the eastern leg (that impacts STEREO) is inside slower solar wind and that part of the ME drives a shock and forms a sheath region (top right panel). On February 23, the shock and sheath impact 1~au, while the HSS is exiting {\it Wind} (bottom left panel). On February 24, the ME impacts both {\it Wind} and STEREO-A with orientations consistent with leg crossings. A ME curvature as drawn in the bottom right panel of Figure~\ref{fig:sketch} can explain how STEREO-A measures a longer-duration ME than {\it Wind}.

While this scenario is overall consistent with the observations, it raises a number of open questions that CME researchers should ponder. The pitch-angle distribution at {\it Wind} and STEREO-A show seemingly different topologies (unidirectional strahl vs.\ BDEs), whereas the expectations are that suprathermal electrons reflect global properties (not local ones) of the ME morphology. As {\it Wind} and STEREO-A magnetic field measurements are consistent with the crossing through the two legs of the ME, it is hard to conceive how the field lines measured at STEREO-A could be closed while the {\em same} field line at {\it Wind} is open. It is possible to reconcile this with our scenario if all magnetic field lines at STEREO-A are closed, some at {\it Wind} are open but originate from the western leg of the ME only and some closed field lines at {\it Wind} having very different intensities along the two legs of  the ME due to the different path lengths.
Additionally, even though it might be less surprising, these measurements emphasize how many of the CME ``properties'' such as the presence/absence of shock, the speed at 1~au, the size, are in fact local properties. This even extends to the presence of sheath regions associated with ME (independently of the presence of a shock). 

Overall, this highlights the need for more multi-spacecraft measurements and dedicated missions. Some additional measurements by two spacecraft will hopefully be possible as STEREO-A comes back to the proximity of Earth in the next two years but measurements by more than two spacecraft and for smaller separations would help further to constrain the CME morphology and properties. Based on the February 2021 event which starts as relatively wide filament but is otherwise a relatively typical CME as seen in remote images, dedicated multi-spacecraft missions would ideally have separations of  $\lesssim 20^\circ$ to maximize the likelihood of having multi-spacecraft measurements (i.e.\ if one spacecraft crosses close to the nose of the ME, another spacecraft at this separation would still observe the same ME).

\begin{acknowledgments}
This research would not have been possible without the longevity and open data policy of STEREO-A and Wind. Research for this work was supported by 80NSSC20K0431 and 80NSSC20K0700. N.~L. acknowledges additional support from 80NSSC20K0197. N.~A. and W.~Y. acknowledge support from AGS1954983 and 80NSSC21K0463.
C.S. acknowledges the NASA Living With a Star Jack Eddy Postdoctoral Fellowship Program, administered by UCAR’s Cooperative Programs for the Advancement of Earth System Science (CPAESS) under award no. NNX16AK22G.
C.~J.~F. acknowledges support from 80NSSC19K1293.
C.~M. thanks the Austrian Science Fund (FWF): P31521-N27, P31659-N27. R.~M.~W. acknowledges support from NASA grant 80NSSC19K0914.
\end{acknowledgments}



\end{document}